# DATE: **D**efense **A**gainst **TE**mperature Side-Channel Attacks in DVFS Enabled MPSoCs

Somdip Dey, *Student Member, IEEE,* Amit K. Singh, *Member, IEEE,* Xiaohang Wang, *Member, IEEE,* and Klaus D. McDonald-Maier, *Senior Member, IEEE*

**Abstract**—Given the constant rise in utilizing embedded devices in daily life, side channels remain a challenge to information flow control and security in such systems. One such important security flaw could be exploited through temperature side-channel attacks, where heat dissipation and propagation from the processing elements are observed over time in order to deduce security flaws. In our proposed methodology, DATE: Defense Against TEmperature side-channel attacks, we propose a novel approach of reducing spatial and temporal thermal gradient, which makes the system more secure against temperature side-channel attacks, and at the same time increases the reliability of the device in terms of lifespan. In this paper, we have also introduced a new metric, Thermal-Security-in-Multi-Processors (TSMP), which is capable of quantifying the security against temperature side-channel attacks on computing systems, and DATE is evaluated to be 139.24% more secure at the most for certain applications than the state-of-the-art, while reducing thermal cycle by 67.42% at the most.

## 1 Introduction and Motivation

As embedded computing systems become more and more ubiquitous, security issues in these and similar systems become more paramount. These embedded systems have to face hostile security threats [1]–[4] such as physical, logical/software-based and side-channel/lateral attacks. Amongst these, side-channel attack is a popular security threat due to ease of access to the physical hardware [1]–[3], where attacks are performed by observing the properties and behavior of the system such as power consumption, thermal dissipation, electromagnetic emission, etc.

Comparatively a lot less documented studies are performed in temperature based side channel attacks [1]–[3] in the embedded computing systems. Due to a rise in the use of heterogeneous multi-processors systems-on-chips (MPSoCs) [5], [6] in the embedded systems and a rise in studies on thermal side channel attacks [3], [7], [8], it is crucial that side channel attacks in such platform should be addressed with utmost importance [8]. Several studies [3], [7], [8] have pointed out processor's core temperature could be used to carry out side channel attacks even when the system has strong spatial and temporal partitioning of resources such as separate partitioning of cache memory, bus bandwidth, etc. But all these studies are focused on general purpose processors (using Complex Instruction Set Computer (CISC) architecture) and not on embedded multi-processors (using Reduced Instruction Set Computer (RISC) architecture), which have different architecture and operating signature from the general purpose ones. On the other hand, we could also see several studies [9], [10] documenting power/energy consumption for different types of instructions. Since there is a direct correlation between power consumption and heat dissipation [11], [12], and armed with the knowledge of power consumption of each instruction executed on processors, it is even easier to launch a side channel attacks on these systems.

In the study [7] thermal covert channels on multi-processors were exploited by observing the exponential rise and fall of CPU temperature while executing a CPU intensive application. In order to verify this behavior on embedded multi-processors we observed the temperature readings on 4 ARM Cortex A-15 big CPU cores on the Exynos 5422 [13] system-on-chip (SoC) while almost idle (the system was only executing background OS processes and the Linux governor was running *ondemand* power scheme), which we denote as *baseline temperature* of the CPU cores. In Fig. 1 we could notice *baseline temperature* of ARM Cortex A-15 big CPUs[1] in both 2D (Fig. 1.(a)) and 3D (Fig. 1.(b)) plots. The reason to provide a visual representation of both 2D and 3D plots is to highlight the spatial as well as temporal thermal gradient because the 2D representation might not be able to provide a microscopic view of spatial thermal gradient over time. The X-axis of the 3D graph represents the temperature, Y-axis of the graph represents time interval and the Z-axis represents the frequency of that particular temperature over time. Since the Odroid XU4 [6] platform, which implements the Exynos 5422 MPSoC, only have temperature sensors on 4 ARM Cortex-A15

- *S. Dey, A. K. Singh and K. D. McDonald-Maier are with the Embedded and Intelligent Systems Laboratory, University of Essex, UK. Corresponding E-mail: somdip.dey@essex.ac.uk*
- *X. Wang is with the School of Software Engineering,South China University of Technology, China.*

1. The graph representing the *baseline temperature* readings for different ARM Cortex A-15 big CPUs closely mimicked the oscillatory pattern that of simple harmonic motion [14].

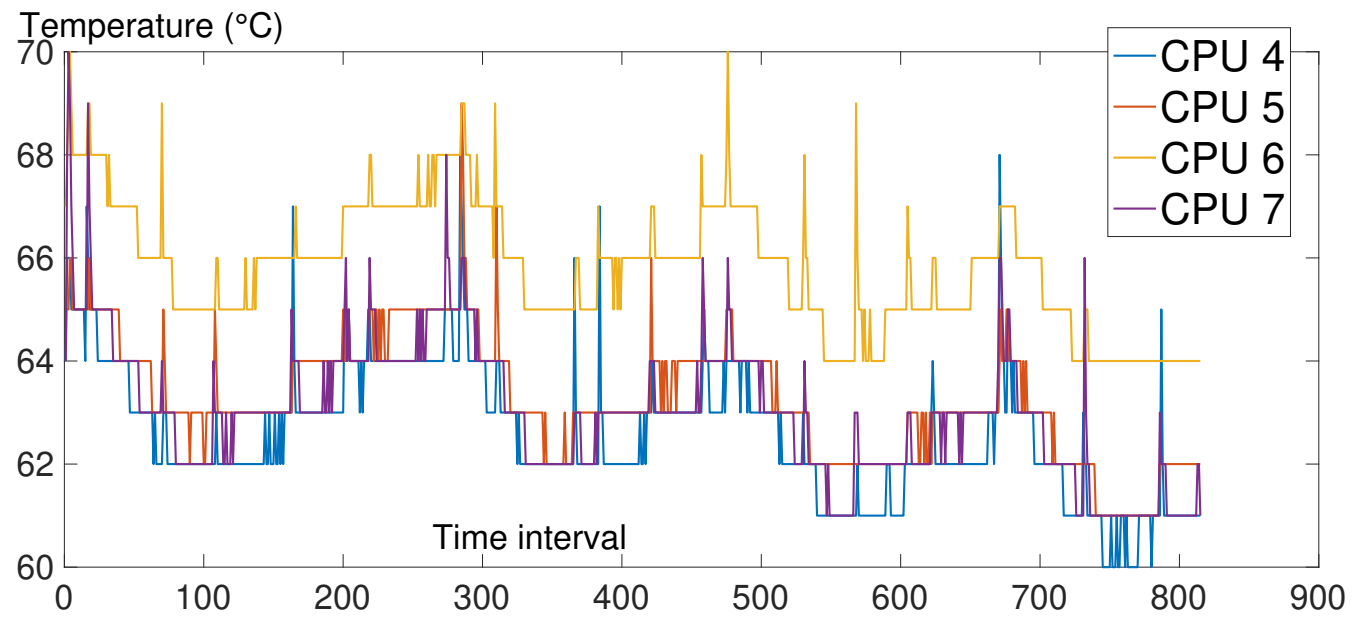

(a) Baseline temperature plot (°C vs Time interval)

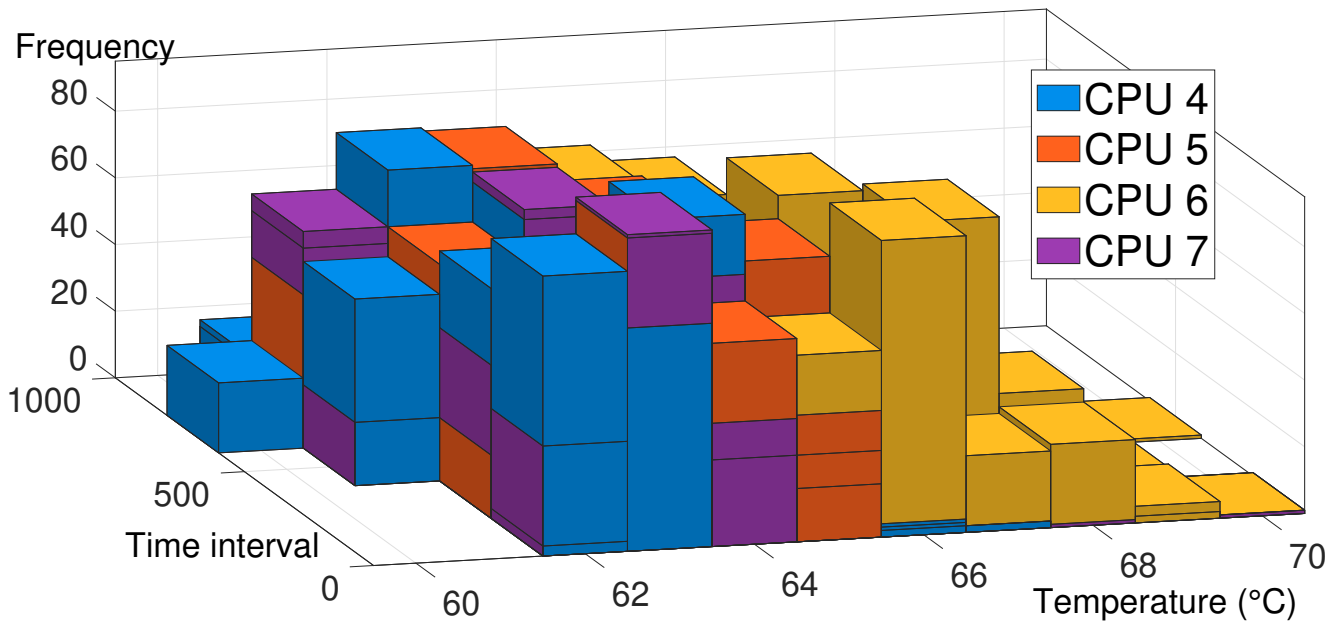

(b) Baseline temperature 3D plot

Fig. 1: Temperature readings on 4 ARM A-15 big CPU cores while idle

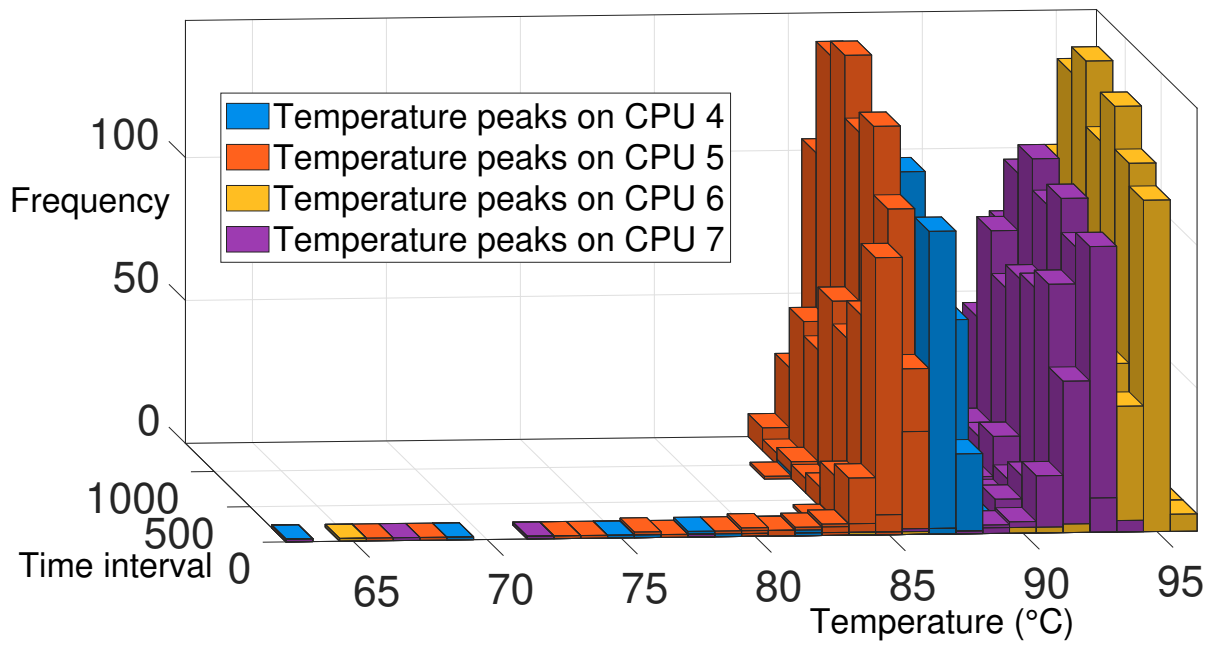

Fig. 2: Temperature peaks on 4 ARM A-15 big CPU cores while executing Blackscholes benchmark

(big) CPU cores, we have only reported on their temperature behavior.

Based on [7] when we executed the Blackscholes benchmark from the PARSEC benchmark suits [15] with default Linux ondemand governor on the Exynos 5422 SoC, we could notice spikes in temperature at certain intervals along with a high frequency in thermal cycles (see Fig. 2). In Fig. 2, we noticed that the temporal and spatial thermal gradient on the CPU cores varies a lot during the execution period. When we compared the thermal behavior of the CPUs during the execution of Blackscholes and *baseline temperature* from our earlier observation, it became evident that heat dissipation from multiple processors could in fact lead to thermal covert channels and hence leading to temperature side-channel attacks.

Most of the relevant published works [3], [7], [8], [11], [12], [16]–[18] in temperature side-channel attacks focus on various different element of pursuing such attack or protect from it. However, none of them has evaluated security against temperature side-channel attack as a cumulative entity. The main reason is that the definition of overall security of a device against temperature side-channel attacks is missing. **Therefore**, in this paper, we define a metric coined as *Thermal-Security-in-Multi-Processors* ($TSMP$), which indicates the cumulative security against such side-channel attacks.

The main objective of this paper is to reduce the value of TSMP for any executing application/task in order to improve the security of the MPSoC from temperature based side-channel attacks. In order to overcome the pressing issue of temperature side-channel attacks, we propose a novel thermal management methodology on Dynamic Voltage Frequency Scaling (DVFS) [5], [19]–[21] enabled MPSoCs in an embedded device, and to the best of our knowledge it is the first documented methodology on securing against thermal side-channel attacks on DVFS. DVFS helps to reduce the energy consumption by executing the workload over extra time at a lower voltage and frequency, which could be accounted for reduced power consumption. We have coined our proposed methodology as $DATE$, Defense Against TEmperature side-channel attacks. Another important advantage of our proposed methodology is that due to the overall reduction of operating temperature the reliability of the device in terms of lifespan also increases. We have also performed temperature based side-channel attack on the latest Samsung Galaxy Note9 [22] (utilizing Exynos 9810 MPSoC [23]) phablet to show the efficacy of $DATE$ by using Covolutional Neural Network (CNN) [24]–[26] based Deep Learning [27] to analyze thermal behavior and predict which password is being used for cryptographic operation. In summary, this paper makes the following contributions.

1) A new metric to quantify the probability of the embedded device prone to temperature based side-channel attacks.
2) A novel thermal management scheme specific to executing application to secure against temperature side-channel attacks and improve reliability of the device in terms of lifespan.
3) Motivational case study with a real attack on real hardware platforms such as Galaxy Note9 [22] and Odroid-XU4 [6].
4) Implementation and validation of our proposed methodology on the real hardware platform (Odroid-XU4) using several benchmark applications.

The rest of the paper is organized as follows. Sec. 2 describes the problem definition and introduces some terms useful to understand the proposed methodology. In Sec. 3, we discuss the proposed methodology along with associated block diagram and algorithms. Sec. 5 explores the different experimentation and evaluation performed to prove the efficacy of our proposed methodology while discussing the future scope of this research. In Sec. 6, we explore the related works on this topic and finally, Sec. 7 concludes the paper.

## 2 Problem Formulation

If we assume that there are $n$ number of processing elements ($p$) in the MPSoC and $P$ represent the set of all the processing elements then we get Eq. 1

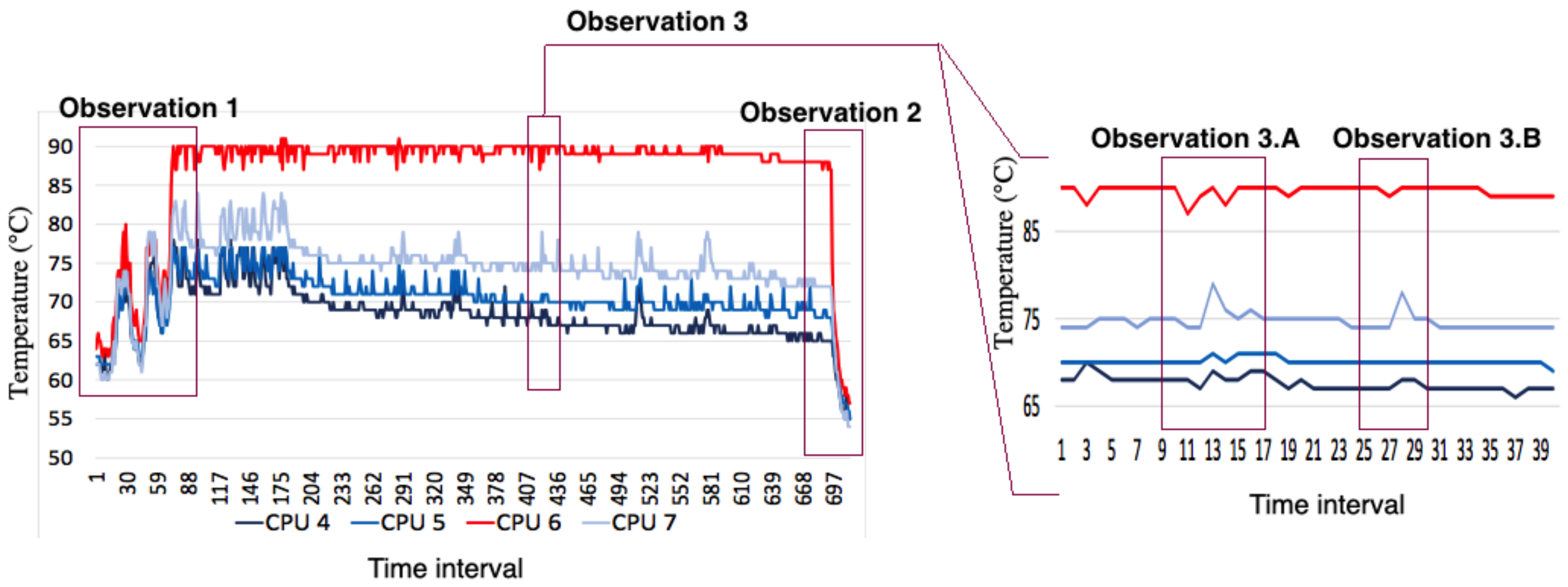

Fig. 3: Observations of thermal behavior of four ARM Cortex A15 (big) CPUs on Odroid XU4 while performing RSA encryption and decryption

$$P = \{p_1, p_2, ....p_n\} \tag{1}$$

Now, if we consider $b_i$ be the *baseline temperature* for a processing element $p_i$, where baseline temperature is the temperature of the corresponding processing element $i$ on the MPSoC while idle, and $t_i$ be the maximum operating temperature of $p_i$ while executing a task $T$, then we get the formula for Baseline Temperature Difference ($BTD$) for $p_i$ for task $T$ as Eq. 2.

$$BTD_{p_i} = \left| t_i - b_i \right| \tag{2}$$

If $S(BTD)$ be the set of all Baseline Temperature Difference for all processing elements for a particular task $T$ in the system such that $S(BTD) = \{ BTD_{p_1}, BTD_{p_2}, ....BTD_{p_n}, \}$, then from the Eq. 2 we can deduce the equation for *Baseline Maximum Thermal Deviation* ($\tau$) as follows:

$$\forall p_i \in P: \ \tau = max(\ S(BTD)\ ) \tag{3}$$

Let's assume that the temperature difference ($d_i$) between any two processing elements such as $t_k$ and $t_j$ be the average operating temperature of $p_k$ and $p_j$, respectively, while executing a task $T$, then we get the following equation:

$$d_i = \left| t_j - t_k \right| \tag{4}$$

Now, for all the processing elements $P$ we will achieve $\frac{n!}{2(n-2)!}$ combinations of temperature difference between any two processing elements in the system. If we consider *Spatial Processing Temperature Differences* ($SPTD$) be a set containing all the combinations ($\frac{n!}{2(n-2)!}$) of temperature differences between the processing elements while executing a task $T$ then we could derive the formula for *Spatial Maximum Thermal Deviation* ($\omega$) as follows:

$$\forall p_i \in P: \ \omega = max(\ SPTD\ ) \tag{5}$$

For quantifying the security of a MPSoC against temperature side-channel attacks, we can derive an equation from the values of $\tau$, $\omega$ and the frequency of thermal cycle during execution of task $T$ ($\theta$). The reason to choose these three variables ($\tau$, $\omega$ & $\theta$) is that from our experiments (described in Section 2.1) we observed that these three variables are directly correlated to thermal behavior of the processing elements. We define the security against temperature side-channel attacks as *Thermal-Security-in-Multi-Processors* ($TSMP$), shown in Eq. 6.

$$TSMP = \frac{1}{\tau + \omega + \theta} \ \Big| \ 0 < TSMP < 1 \tag{6}$$

In Eq. 6 if the value is closer to 1 then it means that the device is more secure against temperature side-channel attacks, whereas, as value closer to 0 means that the device is more prone to security threats related to temperature side-channel. We also have to keep in mind that $(\tau + \omega + \theta)$ can never be $\infty$ or 0 in reality nor the value of $\tau$, $\omega$ and $\theta$ could be decimal as per our platform. Now, from the Eq. 6 in order to improve the security of a device against temperature side-channel attacks our main objective is to keep the total value of $(\tau + \omega + \theta)$ as low as possible such that the value of $TSMP$ is as close to 1 as possible. Therefore, we could define our problem as follows.

**Given**: An application and a MPSoC platform with DVFS support.

**Find**: Maximum value of $TSMP$.

**Subject to**: Meeting performance requirement of each application within available MPSoC resources.

### 2.1 Importance of $\tau$, $\omega$ and $\theta$ in TSMP metric

To understand the key factors that are related to establishing thermal side-channel attacks on a MPSoC such as Odroid XU4 [6] , we performed RSA encryption and decryption several times on the platform and observed the thermal behavior of the four ARM A15 (big) CPU cores. On the Odroid XU4, the 8 CPU cores (big.LITTLE) are denoted as CPU 0, CPU 1 to CPU 7, where CPU 0 to CPU 3 are ARM A7 (LITTLE) cores and CPU 4 to CPU 7 are ARM A15 (big) cores. We are only performing the encryption and decryption on one of the big CPU cores (CPU 6) so that we could observe the thermal behavior on the other neighboring idle CPU cores (CPU 4, 5 and 7). Fig. 3 shows one such instance of the experiments performed where we are observing the thermal behavior 3 seconds before the RSA encryption and decryption is executed till 2 seconds after the execution completes. In Fig. 3 we have highlighted three important observations (see Observation 1, 2 and 3 in the figure), which were common in all the experiments performed. On the Y axis of Fig. 3 we could notice the temperature in degree centigrades of each of the A15 CPU cores, colour coded separately for individual CPU cores, and the X axis represents the time interval with 100 milliseconds gap between each interval. In Observation 1 we could notice that the temperature drastically increases on CPU 6 within 3 seconds and the most important thing to notice here is that not only the temperature of CPU 6 increases but at the same time temperature of all neighboring A15 CPU cores (CPU 4, 5 and 7) also increase. The reason for this phenomenon is due to heat dissipation and propagation in the lateral direction. In Observation 2 we can notice that once the encryption and decryption complete on CPU 6 the temperature plummets not only on that core but also on the neighboring cores. Although in this observation it could be noticed that temperature plummets over 2 seconds and not instantaneously. As stated by Masti et al. [7], thermal covert channels can be exploited by observing the exponential rise and fall of CPU temperature. Therefore, Observation 1 and Observation 2 proves the theory proposed in the study [7]. More importantly, *Baseline Maximum Thermal Deviation* ($\tau$) and *Spatial Maximum Thermal Deviation* ($\omega$) directly reflect the issues mentioned in Observation 1 and 2 of Fig. 3.

More interestingly, if we notice Observation 3 (3.A and 3.B) we can find that whenever there is a spike in temperature on CPU 6 there is also a spike in temperature on the neighboring CPUs due to heat propagation. Thus proving that when the temperature of one CPU core increases, due to heat propagation from that CPU core to the nearby CPU cores the thermal behavior of these adjacent cores are also affected. Therefore, potentially opening up a vulnerability if one such adjacent CPU core could be used to establish a temperature side-channel attack.

Since, $\tau$ and $\omega$ reflect the temporal as well as spatial thermal gradient difference of the CPU cores while they are executing a task (or set of tasks) and being idle, the aforementioned variables are very important in understanding the vulnerability of MPSoCs against temperature side-channel attack. Whereas, the frequency of thermal cycle during execution of a task (or set of tasks) ($\theta$) directly reflects the phenomenon of Observation 3 (3.A and 3.B) and hence $\theta$ also plays an important role in deducing vulnerability related to temperature side-channel attack. Therefore, it is crucial to define the *Thermal-Security-in-Multi-Processors* (*TSMP*) metric incorporating $\tau$, $\omega$ and $\theta$.

# 3 Proposed Methodology: DATE

## 3.1 Overview of DATE

An overview of the proposed DATE methodology is illustrated in Fig. 4. It has some offline and online steps. During offline profiling computed results for various applications are used to identify the appropriate temperature ($t_{App_i}$) and frequency ($f_{App_i}$) such that a cumulative least value for ($\tau+\omega+\theta$) (see Eq. 6) is achieved. Since for each application we have to explore all the possible frequency values along with appropriate temperature and performance threshold such that TSMP is maximum for $App_i$, we do the profiling offline as part of the design space exploration. Once we have found out the best possible frequency and temperature which yields the least value for ($\tau+\omega+\theta$), we save the frequency and temperature values along with some other parameters. These values are later used during the online step where we set the frequency and temperature threshold of the system to these saved values. In DATE, we call the offline step as *Learning Module* whereas, the online step is called as *Decision Module*. Since, operating temperature of the device directly affects the reliability of such device in terms of lifespan, hence, DATE methodology affects the reliability of the device in a positive manner (more details in Sec. 5.3).

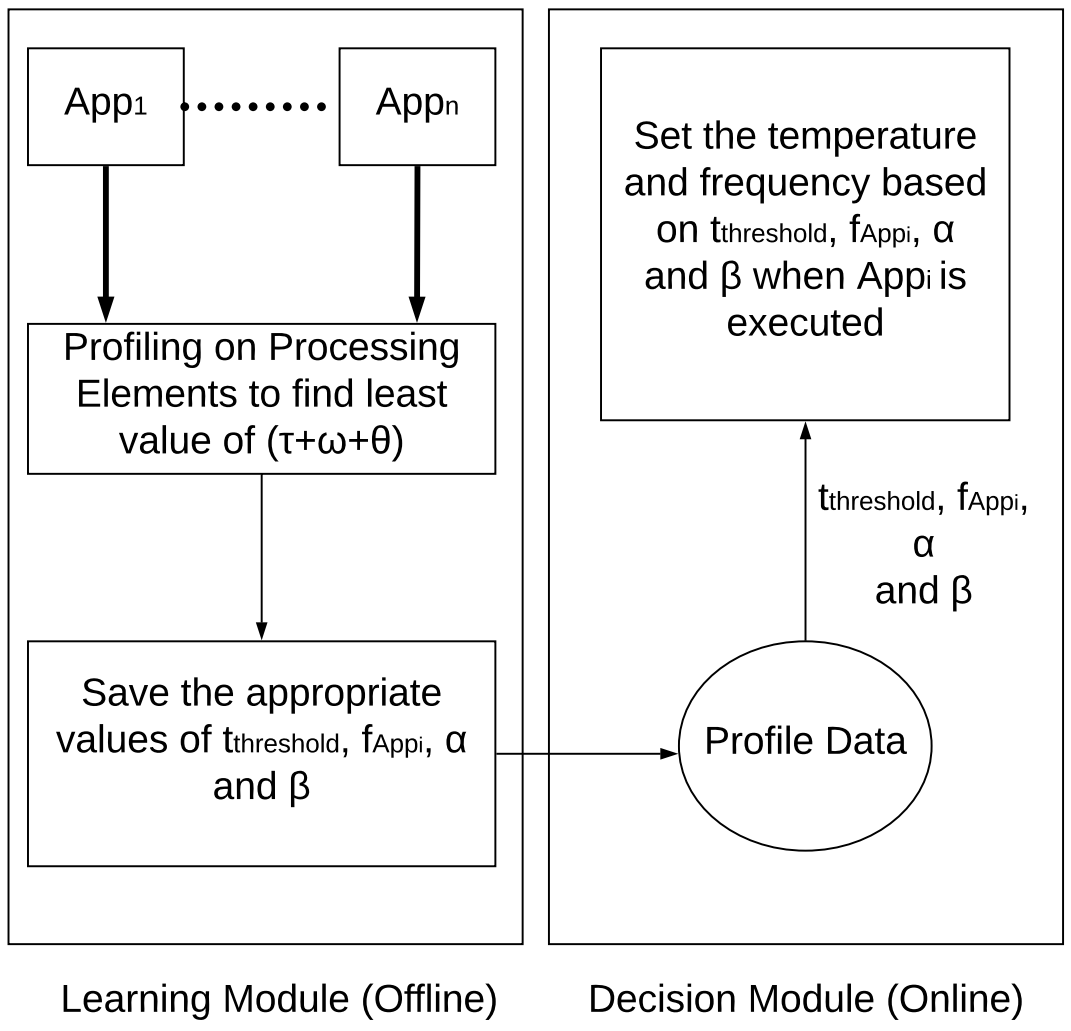

Fig. 4: Proposed Methodology: DATE

## 3.2 Learning Module

The complete algorithm explaining the working of *Learning Module* is provided in Algo. 1.

In the study [11] through experiments on real devices, it was noticed that on MPSoCs power and temperature follow a relationship of quadratic function. On the other hand, in DVFS enabled MPSoCs dynamic power and frequency follow a quadratic relationship due to the power consumption ($P \propto V^2 f$) being directly affected by the operating frequency and voltage [28]–[30]. However, to make our approach fast during run time (*Decision Module*) we assume that frequency and temperature follows a linear relationship represented by Eq. 7.

$$t_i = \alpha \times f_i + \beta \tag{7}$$

In Eq. 7, $f_i$ is the operating frequency at time instance $i$, $t_i$ is the operating temperature at the same time instance, $\alpha$ is the relationship variable and $\beta$ is the intercept variable.

Now, we first set a performance deadline (for our case we choose execution time deadline for each application) and then start executing $App_i$ and vary the frequency as well as the temperature threshold of the system to observe the performance. Here, the temperature threshold is the operating thermal cap, which can not be surpassed by the system during the execution of the application. If we consider $t_{available}$ is the set of all available operating temperature ($t_{available} = \{t_1, t_2, \ldots t_i\}$) of the system, which will be used as the thermal cap during the profiling, then for each instance of operating temperature ($t_{available_i}$) from $t_{available}$ we monitor the performance while setting the temperature threshold to $t_{available_i}$.

While monitoring performance the frequency of the PEs is reduced by one frequency scaling step (we start at the maximum frequency and this is denoted as *ReduceFrequencyByOneStep()* in Algo. 1) and returns the reduced frequency instance ($f_i$). We monitor the temperature of each PE (denoted as *MeasureTemperatureValue()* in Algo. 1) and returns the set of temperature values of all the PEs during the profiling. If the performance deadline for the application is met then the maximum value among the set of temperature is used as the temperature threshold. By this approach we leverage DVFS capability of the system and modify the frequency till the performance constraint is met and try to achieve least possible values for $\tau$ , $\omega$ and $\theta$. After we have found the appropriate value of system's temperature threshold/cap ($t_{App_i}$) and frequency ($f_{App_i}$) we note the value of $\alpha$ and $\beta$ (from Eq. 7). For $App_i$ the relationship variable and intercept are denoted by $\alpha_{App_i}$ and $\beta_{App_i}$ respectively. For each profiled application we only save the values of maximum operating temperature threshold/cap ($t_{threshold_{App_i}}$) which should not be surpassed during the execution of $App_i$, operating frequency ($f_{App_i}$), $\alpha_{App_i}$, $\beta_{App_i}$ and baseline temperature ($b_{App_i}$) during the profiling period on the memory to be used in the *Decision Module*. Here, the baseline temperature ($b_{App_i}$) for the profiling application is the same baseline temperature, which is recorded before and after the application is executed such that $b_{App_i} = b_i$ in Eq. 2. In practice, we actually save the value of $b_{App_i}$ for each processing element in the system as a set of baseline temperatures ($B_{App_i}$).

**Algorithm 1:** Learning Module Execution

**Input:**
1. $Perf_{max}$: performance deadline for an application instance $App_i$
2. $B_i$: a set of baseline temperature ($b_i$) for each processing element (PE)

**Output:** $< t_{threshold_{App_i}}, f_{App_i}, \alpha_{App_i}, \beta_{App_i}, B_{App_i} >$
**Initialize:** $\tau + \omega + \theta = \infty$;

**foreach** $t_{available_i}$ *in* $t_{available}$ *of the system* **do**
  $t_{threshold_{App_i}} = t_{available_i}$;
  %$Perf_{App_i}$ is the current performance of the application%
  **while** $Perf_{App_i} \leq Perf_{max}$ **do**
    $f_i$ = ReduceFrequencyByOneStep(); (see Sec. 3.2)
    $T_i$ = MeasureTemperatureValue();
    $\tau_i$ = CalculateTau($B_i$); %see Eq. 3%
    $\omega_i$ = CalculateOmega($B_i$); %see Eq. 5%
    $\theta_i$ = CalculateTheta(); %see Sec. 3.2%
    **if** $(\tau_i + \omega_i + \theta_i) < (\tau + \omega + \theta)$ **then**
      $\tau = \tau_i,\ \omega = \omega_i,\ \theta = \theta_i$;
      $t_{threshold_{App_i}} = max(T_i),\ f_{App_i} = f_i$;
  $< \alpha_{App_i}, \beta_{App_i} >$ = CalculateAlphaBeta($t_{threshold_{App_i}}$, $f_{App_i}$); (Sec. 3.2)

return $< t_{threshold_{App_i}}, f_{App_i}, \alpha_{App_i}, \beta_{App_i}, B_{App_i} = B_i >$;

### 3.3 Decision Module

The algorithm for Decision Module is provided in Algo. 2. In the *Decision Module*, a software agent, which we call as *DATE* as well, first loads the baseline temperature ($b_{App_i}$) for the profiled application and compares the current temperature with baseline temperature during this time instance ($b_{now}$). The set of baseline temperatures for all PEs is $B_{now}$. Since baseline temperature can vary depending on the surrounding and other unforeseen circumstances, it is important DATE is aware of the difference between the baseline temperature when the application was profiled and the baseline temperature for the time period when the Decision Module of DATE acts and could be represented as $b_{diff} = b_{App_i} - b_{now}$. For set of PEs it is $B_{diff}$.

Now, if we assume that the value of $\alpha$ and $\beta$ remains almost same (subject to some obvious deviation and we refer to it as error) then from Eq. 7 we can get the equation (see Eq. 8) that could define the relationship between two different operating temperature and their associated frequencies.

$$f_2 = \frac{(t_2 - t_1)}{\alpha} + f_1 \tag{8}$$

We can modify the Eq. 8 to reflect the governing equation (see Eq. 9) to decide the operating temperature threshold and associated frequency by DATE.

$$f_{desired} = \frac{(t_{desired\ threshold} - t_{threshold_{App_i}})}{\alpha} + f_{App_i} \tag{9}$$

If the value of $b_{diff}$ is negative then DATE reduces the operating frequency by evaluating the new operating frequency value while considering that $b_{diff} = (t_{desired\ threshold} - t_{threshold_{App_i}})$ in Eq. 9. But if the value of $b_{diff}$ is positive then Eq. 9 could be modified as follows:

$$f_{desired} = \frac{(t_{desired\ threshold} + b_{diff})}{\alpha} + f_{App_i} \tag{10}$$

If the value of calculated $f_{desired}$ is more than the available frequency of the CPU cores then DATE do not modify the operating frequency, thus, $f_{desried} = f_{App_i}$ but modifies the operating temperature threshold as $t_{desired\ threshold} = b_{diff} + t_{threshold_{App_i}}$. While executing the application, if the operating temperature of the CPU cores tries to go higher than the operating temperature threshold ($t_{desired\ threshold}$) then DATE drops the frequency ($f_{desired}$) to next frequency scaling level, which is again evaluated from the Eq. 8 by assuming that the operating temperature of the system needs to be reduced by 1° Centigrade. The governing equation to achieve this is as follows:

$$f_{desired} = f_{now} - \frac{1}{\alpha} \\ where,\ t_{desired} - t_{now} = -1 \tag{11}$$

## 4 Case Study: Performing a real attack on Galaxy Note9 and Odroid XU4

We performed a live attack on the Galaxy Note9 mobile device, which is one of the top rated smart phones of 2018 and 2019 [31], [32]. To prove the scalability and affect of the attack we also executed the same methodology on Odroid XU4. We executed a program sitting on one of the LITTLE cores on the device, while snooping the thermal behavior of the big core cluster during several encryption (AES-256 [33]) operations

**Algorithm 2:** Decision Module Execution

**Input:**
$< t_{threshold_{App_i}}, f_{App_i}, \alpha_{App_i}, \beta_{App_i}, B_{App_i} >$ & $B_{now}$
$B_{diff}$ = CalculateBaselineDiff($B_{App_i}$, $B_{now}$); (Sec. 3.3)
**foreach** *PE ($P_i$) in the set of PEs (P) on the system* **do**
    $b_{PE_{diff}}$ = GetBaselineDiff($B_{diff}$, $P_i$); %Function to get the baseline difference ($b_{PE_{diff}}$) for $P_i$
    **if** $b_{PE_{diff}} > 0$ **then**
        $f_{desired}$ = CalculateDesiredFrequency($b_{PE_{diff}}$); (using Eq. 10)
        SetFrequencyOfPE($f_{desired}$); %Function to set the frequency of the PE with the value of $f_{desired}$%
        SetTempThres($t_{threshold_{App_i}}$); %Function to set the maximum thermal cap of the PE as $t_{threshold_{App_i}}$%
    **else**
        $f_{desired}$ = CalculateDesiredFrequency($b_{PE_{diff}}$); (using Eq. 11)
        %$f_{PE_{max}}$ is the maximum frequency of the PE%
        **if** $f_{desired} > f_{PE_{max}}$ **then**
            $f_{desired} = f_{App_i}$; SetFrequencyOfPE($f_{desired}$);
            $t_{desired\ threshold} = b_{PE_{diff}} + t_{threshold_{App_i}}$; (see Eq. 10 and Eq. 11)
            SetTempThres($t_{desired\ threshold}$);
        **else**
            SetFrequencyOfPE($f_{desired}$);
            SetTempThres($t_{threshold_{App_i}}$);

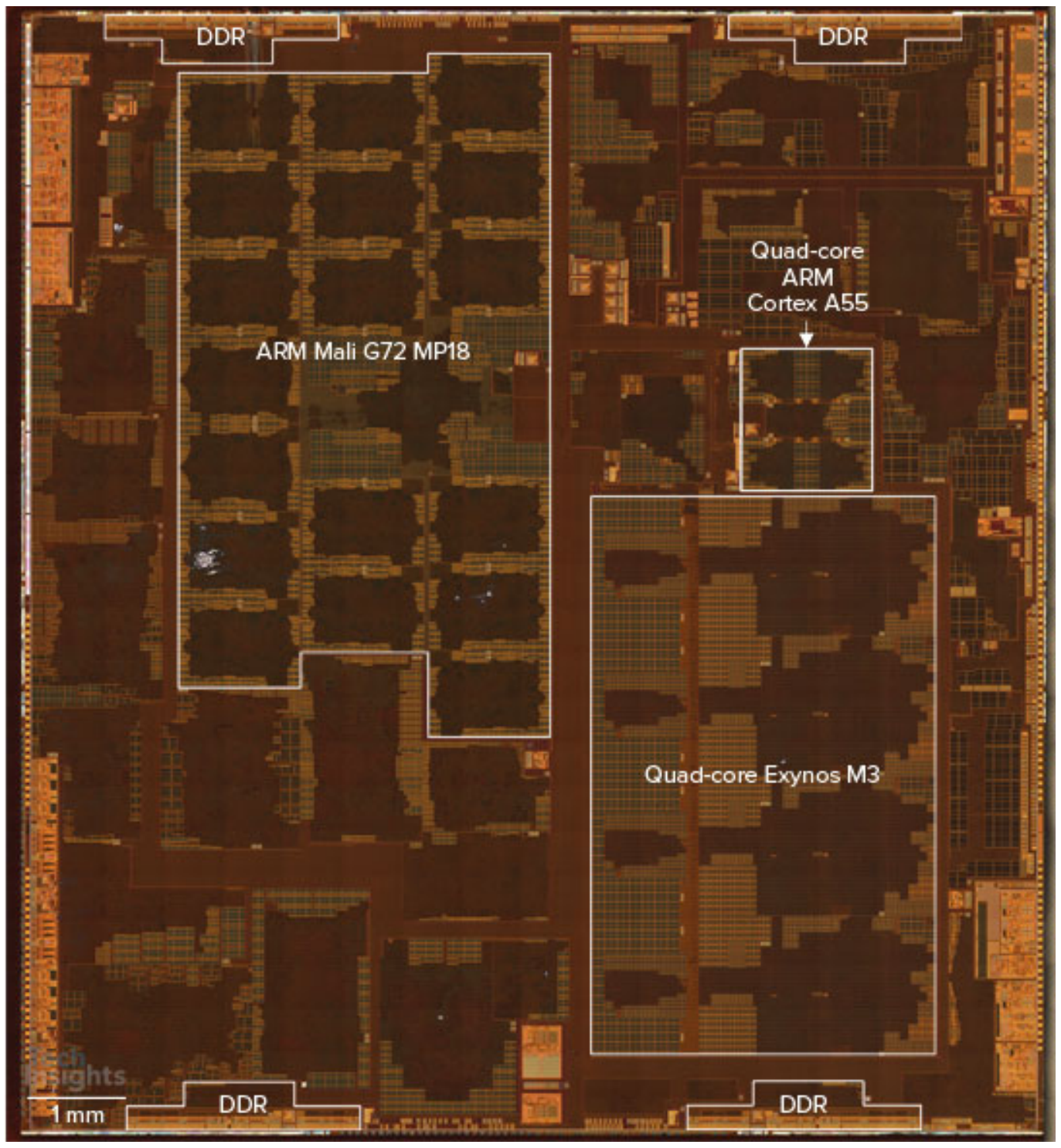

Fig. 5: Exynos 9810 [23] MPSoC block diagram highlighting major components

of a text file with 4 of the most common passwords used by the users. The main motive of this attack is to evaluate the vulnerability of the latest smart-phone against thermal side-channel attack. More details on the device, experimental setup and evaluation of the attack is provided in the following subsections.

### 4.1 Hardware & Software Infrastructure of Note9

Nowadays heterogeneous MPSoCs consist of different types of cores, either having the same or different instruction set architecture (ISA). Moreover, the number of cores of each type of ISA can vary based on MPSoCs and are usually clustered if the types of cores are similar. For this research, we have chosen an Asymmetric Multi-processors (AMPs) system-on-chip (AMPSoC), which has clustered cores on the system. We chose to execute a real attack on Galaxy Note9 [22], which is the latest mobile device from Samsung and utilizes the Exynos 9810 MPSoC [23]. The block diagram of Exynos 9810 is provided in Fig. 5. Exynos 9810 has two CPU clusters, one for big CPU cores consisting of 4 Mongoose 3 CPU cores, and the other cluster for LITTLE CPU cores consisting of 4 Cortex A-55 CPU cores. The Mongoose 3 CPU cores allow cluster wise DVFS and has 18 frequency scaling levels ranging from 650 MHz to 2704 MHz (2704 MHz, 2652 MHz, 2496 MHz, 2314 MHz, 2106 MHz, 2002 MHz, 1924 MHz, 1794 MHz, 1690 MHz, 1586 MHz, 1469 MHz, 1261 MHz, 1170 MHz, 1066 MHz, 962 MHz, 858 MHz, 741 MHz, 650 MHz). Whereas, the LITTLE Cortex-A55 CPU cores has 10 frequency scaling levels ranging from 455 MHz to 1794 MHz (1794 MHz, 1690 MHz, 1456 MHz, 1248 MHz, 1053 MHz, 949 MHz, 832 MHz, 715 MHz, 598 MHz, and 455 MHz).

The Galaxy Note9 was running on Android 9 (Pie) [34] OS utilizing Linux kernel version 4.9.59, which has only one governor named *schedutil* based on Energy Aware Scheduling (EAS) [35]. The Galaxy Note9 is only utilized for this case study to prove the threat of thermal side channel attack in real popular commercial device.

### 4.2 Hardware & Software Infrastructure of Odroid XU4

We also chose Odroid XU4 board [6] (see Fig. 6) to execute the attack in order to verify the affect and scalability of thermal side-channel exploitation on a device other than Galaxy Note9. We also used Odroid XU4 for the rest of the experimentation and validation because of more availability of software execution support since Galaxy Note9 is vendor locked to only modify certain portions of the Android OS. Odroid XU4 employs the Samsung Exynos 5422 [13] MPSoC. As mentioned earlier Exynos 5422 MPSoC contains clusters of big and LITTLE cores. This MPSoC provides DVFS feature per cluster, where the big core cluster has 19 frequency scaling levels, ranging from 200 MHz to 2000 MHz with each step of 100 MHz and the LITTLE cluster has 13 frequency scaling levels, ranging from 200 MHz to 1400 MHz, with each step of 100 MHz.

The Odroid XU4 was running on UbuntuMate version 14.04 (Linux Odroid Kernel: 3.10.105), which was running on performance governor.

### 4.3 Dataset and CNN Model

To perform the attack we chose 4 of the 25 most common passwords of 2017 and 2018 [36], [37] as surveyed by the Internet security firm SplashData. The 4 common passwords used by the user, which are chosen for our attack, are *123456, passw0rd, 111111* and *football*. We executed AES-256 [33] encryption on a text file using the aforementioned passwords 500 times for each password. The reason to choose AES-256 is because of its popularity. The encryption operations were performed on the 4 Mongoose 3 big CPU cores (big CPU cluster) while one of the LITTLE cores snoop the operating temperature data of the big CPU cluster. This is due to the fact that only one thermal sensor is present on the big CPU cluster of Galaxy Note9 rather than individually placed on each big CPU core. After the temperature data for each password were collected, we transformed the data points into a graphical representation in order to be fed to a pre-trained CNN [38] for training and classification purposes. Since on the Odroid XU4 thermal sensors are individually placed on each big CPU cores, we chose to only record the thermal behavior of one of the big CPUs (CPU 7) for this experimental attack.

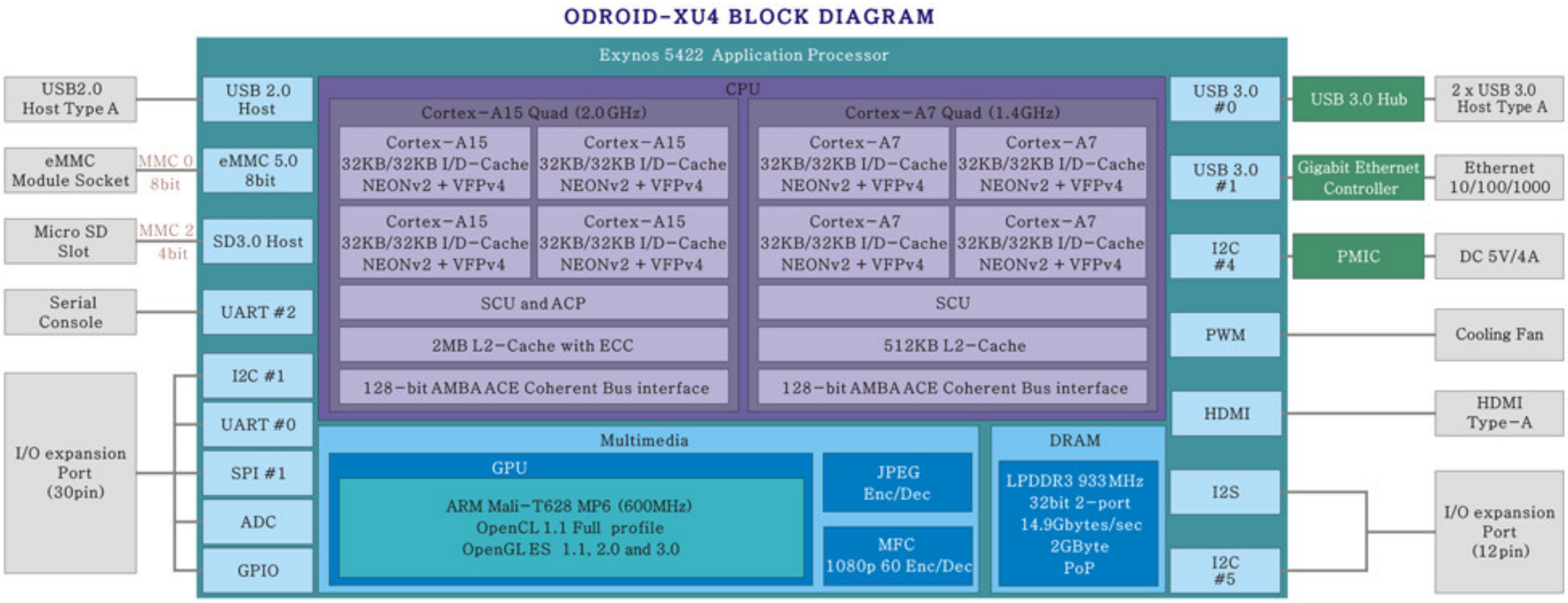

Fig. 6: Odroid XU4 [6] MPSoC block diagram highlighting major components

We chose ResNet50 CNN [26] as our network model, which could be used to train our graphical thermal data using *Transfer Learning* [38]. Since, CNN based Deep Learning usually requires a lot of training data in order to be used successfully for classification purposes, the network models are trained on large dataset such as ImageNet [39], consisting of millions of images, and then the weights and parameters of the model are saved and used later to train for a specific target application. This way of training is called *transfer learning* and we have used ResNet 50, which is pre-trained on ImageNet. Since, ResNet50 is pre-trained on ImageNet and has the last fully connected layer to classify 1000 different classes/labels from the ImageNet, we had to modify the last fully connected layer to suit our target application.

Since, we have 4 passwords as the classes/labels to classify, we removed the last fully connected layer, which was pre-trained on ImageNet, to cater for only 4 labels instead. We have also added a new fully connected layer (called Dense in Fig. 7), dropout regularization (called Dropout in Fig. 7) and a softmax function along with a classifier (called PREDICTIONS in Fig. 7) to only predict for 4 passwords (labels). In Fig. 7 we have shown the modified fully connected layer along with the custom classifier used in our ResNet50 CNN model. Here, the dropout regularization is used for better generalization and the softmax function is used for probability distribution for the 4 labels/classes. In Fig. 8, we show the graphical representation of the thermal behavior for each password label, which is fed

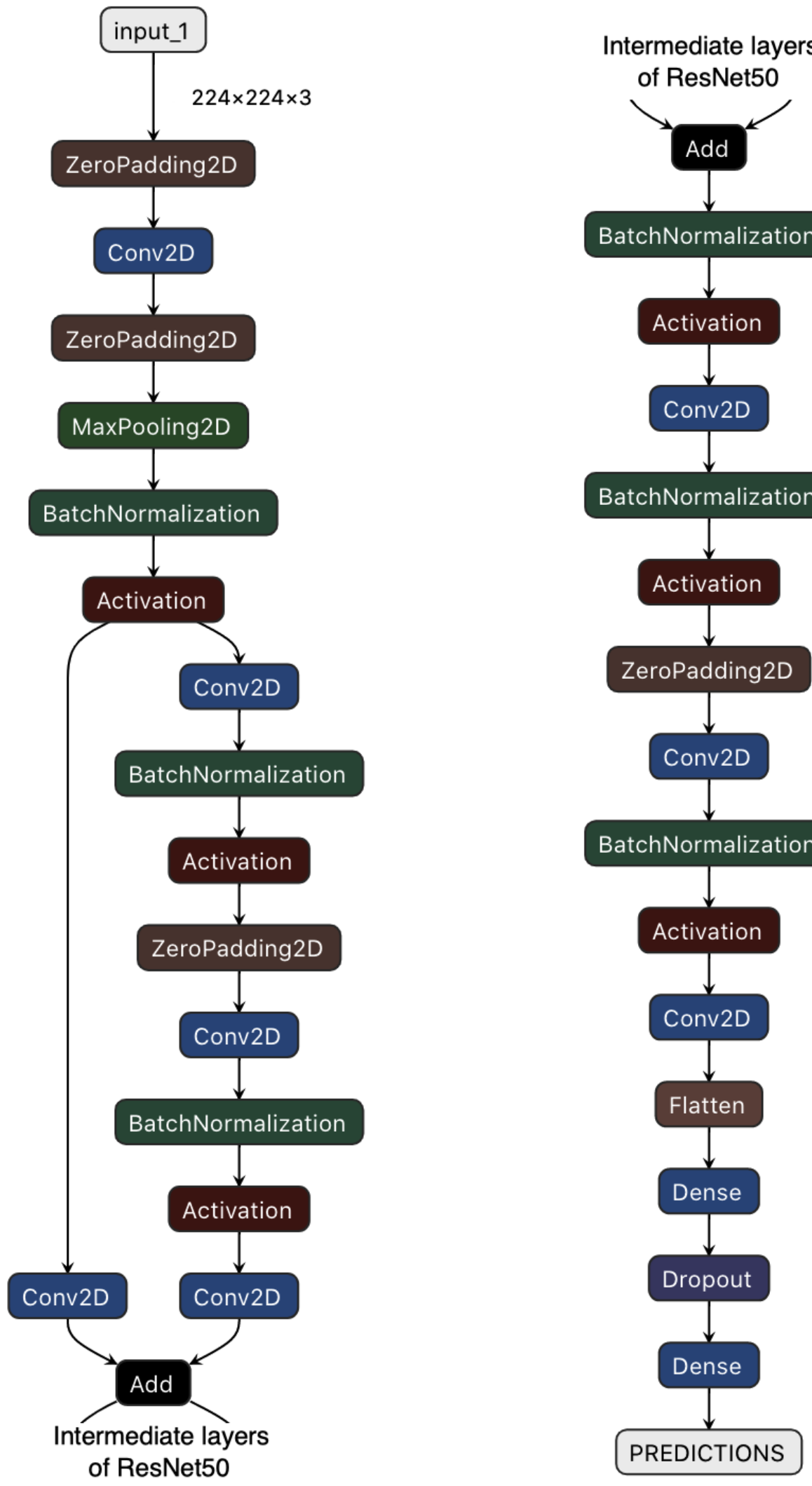

Fig. 7: Architecture of ResNet50 CNN model with modified last fully connected layer used in password classification

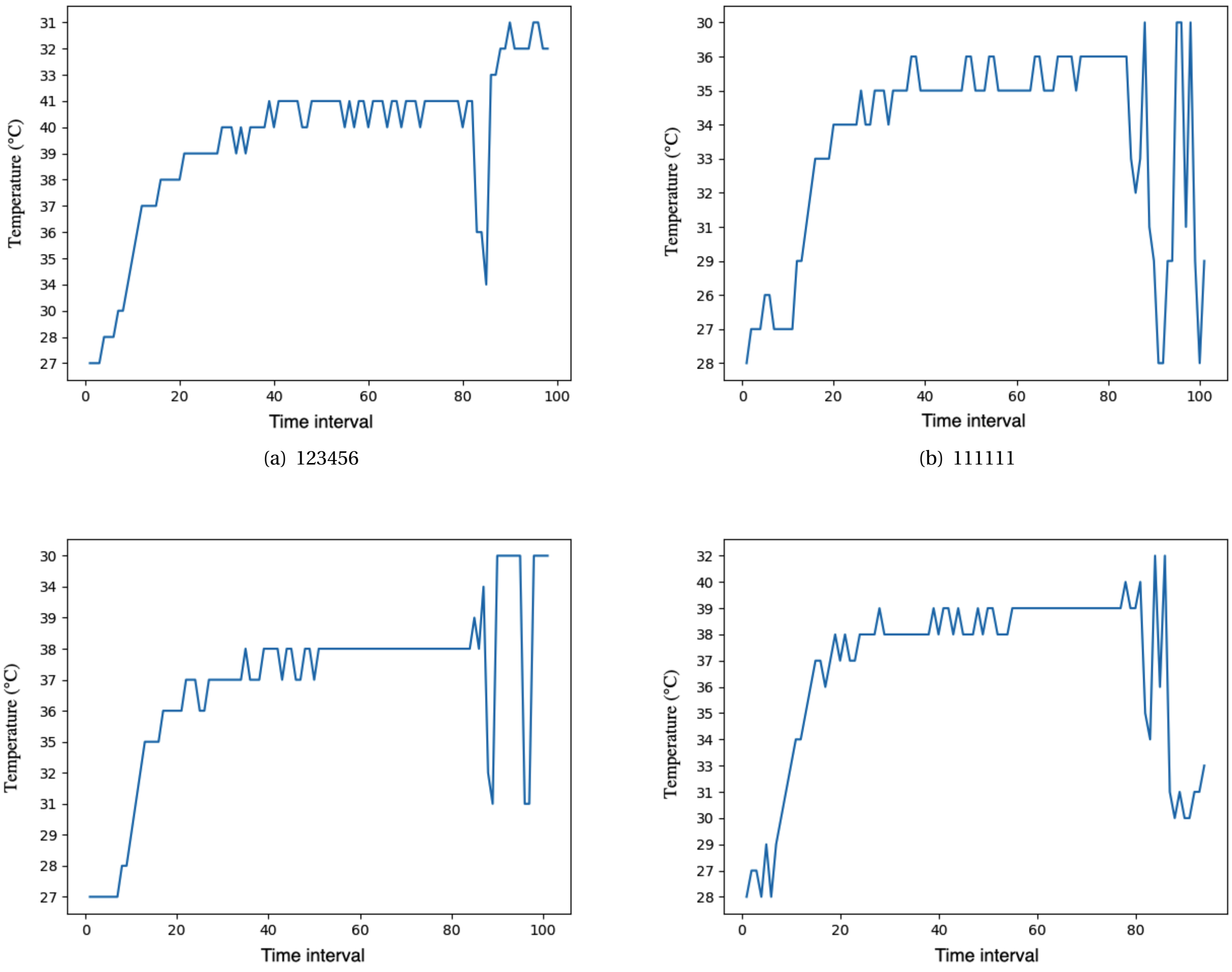

(a) 123456

(b) 111111

(c) passw0rd

(d) football

Fig. 8: Graphical representation of thermal behavior (time interval vs temperature in °C) of encryption operation using the following passwords: *111111, 123456, passw0rd & football*

to the ResNet50 CNN model for training and classification purpose.

From the 500 graphical data for each password label, we separated 100 graphical data for cross-validation testing purpose. Whereas, 75% of the remaining 400 graphical data for each password label were used for training and rest of the 25% is used for validation during the training period. Validation data is used to provide an unbiased evaluation of a model fit on the training dataset while tuning hyperparameters of the model.

**Code availability**: Program codes to implement the attack and generate the dataset could be accessed from ## Codewillbeprovideduponacceptance##.

### 4.4 Experimental Results

After training the ResNet50 model with the training and validation dataset from Note9, the total prediction accuracy achieved is 36.50%. When we used 400 graphical data (100 data from each password label) for cross-validation to evaluate the accuracy of the trained ResNet50 CNN in predicting the password being used for encryption operation, the CNN could predict 146 instances correctly and achieving a prediction accuracy of 36.50%. Whereas, when we implemented DATE on the system to improve the security against thermal side-channel on Note9 and performed the same training methodology with the CNN, the prediction accuracy achieved by the CNN model after the training is 26.75%. When we used the 400 graphical data for cross-validation the CNN was only able to successfully predict 25.25 % (101 out of 400 data). Hence, using DATE to secure the system against thermal side-channel attack such as the one performed in this section, we are able to achieve 9.75% (for training process) and 11.25% (for cross-validation) increase in security.

When we performed the same training on the thermal data collected from Odroid XU4, the CNN achieved a training prediction accuracy of 56.75% when DATE was not implemented on the system. The CNN was also able to predict 52% of the 400 graphical data (208 out of 400) during the cross-validation. After we implemented DATE on the Odroid XU4 the training prediction accuracy achieved by the CNN is 26.75% and was able to predict 29.5% of the 400 data (118 out of 400) during cross-validation. Therefore, using DATE we are able to achieve an increase in security against thermal side-channel by 30% during training period and an increase in security by 22.5% during cross-validation.

In order to determine whether our CNN is classifying the thermal data based on the features of the thermal peaks, we utilized Gradient-weighted Class Activation Mapping (Grad-CAM) [40] to visualize in which areas of the graphical data the CNN was focusing on to predict which password is being used for encryption process. Grad-CAM methodology uses the gradient of the target classes flowing into the last convolutional layer to produce a coarse localization map highlighting the important regions in the image (graphical data) for predicting the class label. Fig. 9 shows the thermal graphical data generated while encrypting the plain text using 111111 as password (Fig. 9.(a)) and the subsequent localization mapping created from the last convolutional layer (Fig. 9.(b)) and then portraying the region of interest as thermal map on the original image (Fig. 9.(c)). Fig. 10, 11 and 12 consequently show the Grad-CAM generated for thermal behavior of the big CPU for individual password's (123456, football and passw0rd) encryption process. From the Grad-cam representation it is evident that our CNN is actually looking at certain thermal peaks to determine which password might have been used for encryption and hence, predict the label accordingly.

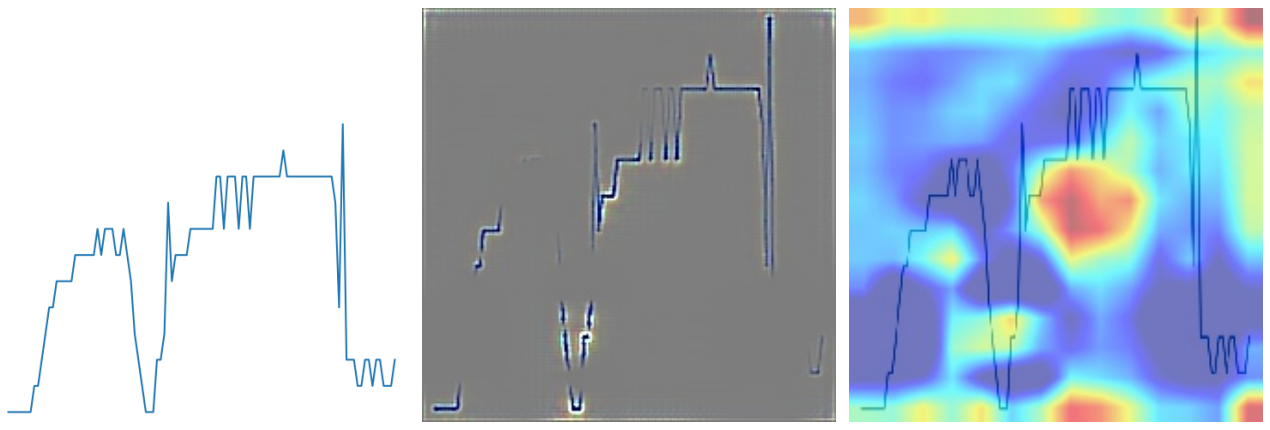

(a) (b) (c)

Fig. 9: Focus area of ResNet50 network on a representative graph of password: 111111

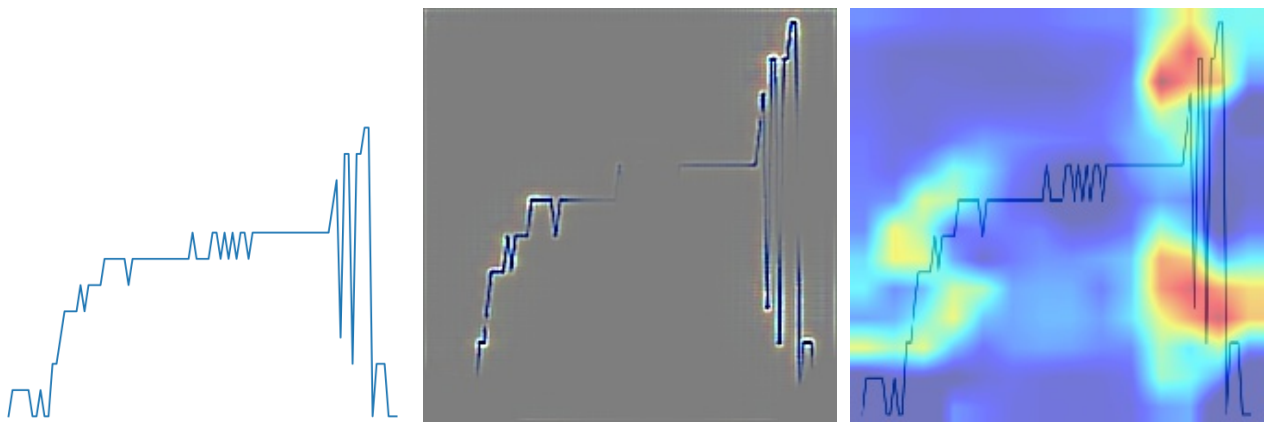

(a) (b) (c)

Fig. 10: Focus area of ResNet50 network on a representative graph of password: 123456

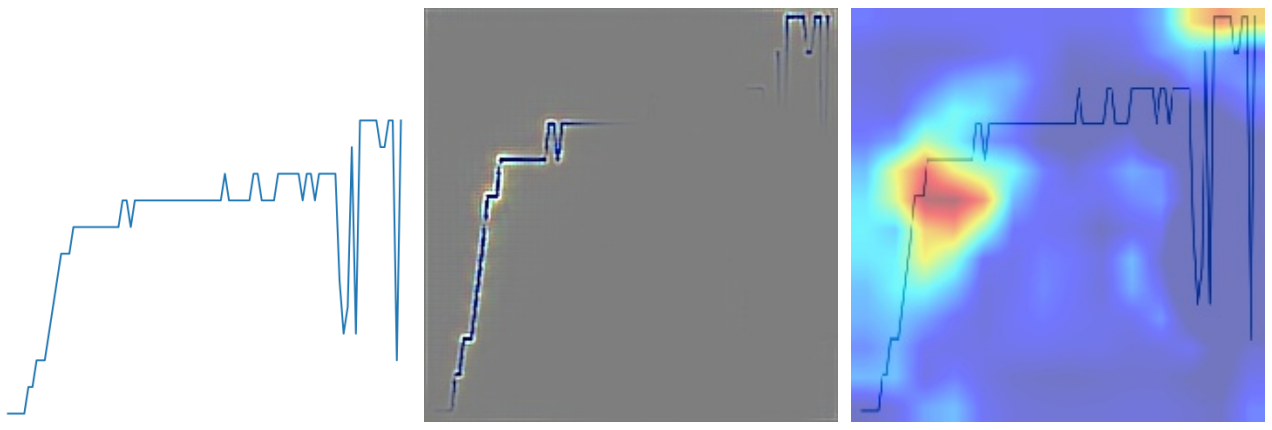

(a) (b) (c)

Fig. 11: Focus area of ResNet50 network on a representative graph of password: Football

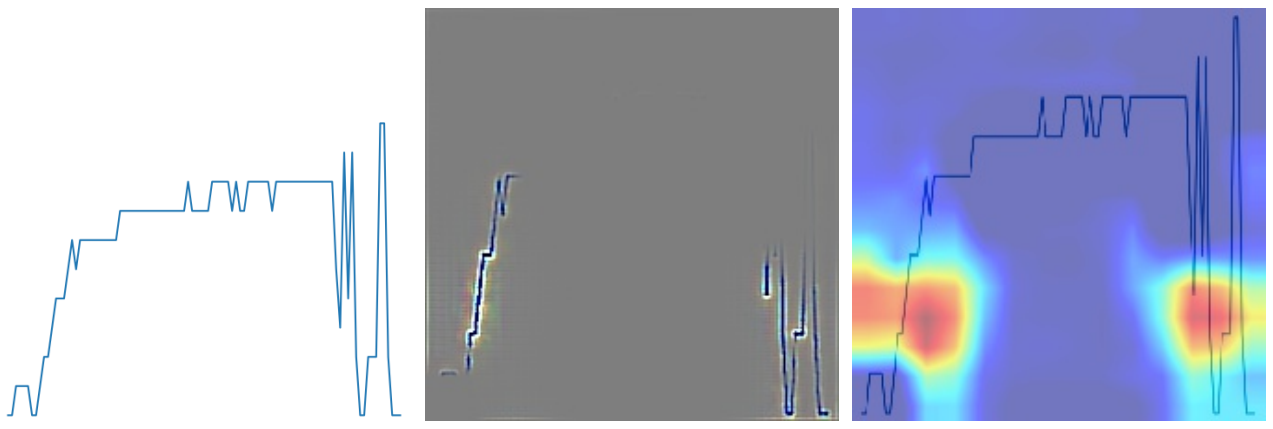

(a) (b) (c)

Fig. 12: Focus area of ResNet50 network on a representative graph of password: passw0rd

# 5 Experimental and Evaluation Results

## 5.1 Hardware & Software Setup for Experiments

Odroid XU4 could be utilized to run on both Android and Linux operating systems, and hence, we chose to perform most of our experiments on Odroid XU4 running on Linux OS. Further, since Linux is more versatile OS than Android and provides the ability to execute several established benhcmark suits such as PARSEC [15] unlike Android, we chose to use Linux for the experimentations provided in this section to show the efficacy and scalability of DATE.

For multi-core systems, multi-threaded applications are heavily used in recent times to represent workloads as they could leverage concurrency and parallel processing. Examples of such applications are available in several benchmarks such as PARSEC [15]. For our experiments we have tried several applications from the PARSEC benchmark such as Streamcluster, Blackscholes, Facesim etc. but to evaluate the effectiveness of our $DATE$ mechanism we chose Blackscholes and Streamclsuter with $native$ option because it closely represented a real-world mixed (compute and memory intensive) workload application and the execution period was long enough to observe the thermal behavior in the system. We also evaluated our approach for other real-world workload such as playing a Youtube video on the Chromium browser and RSA [41] encryption and decryption algorithm for 512, 1024, 2048 and 4096 bits. We have run all our experiments on UbuntuMate version 14.04 (Linux Odroid Kernel: 3.10.105).

## 5.2 Experimental Results

In Fig. 13, 14 and 15 we can see step by step evaluation of Baseline Maximum Thermal Deviation ($\tau$), Spatial Maximum Thermal Deviation ($\omega$) and frequency of thermal cycle ($\theta$) during execution of the application for different applications using DATE. In Fig. 13 Base T. represents the baseline temperature for different ARM Cortex A-15 CPUs.

Table 1 shows TSMP values for different applications when various resource management methodologies are employed. . We evaluated DATE against ondemand and performance governor of Linux and the state-of-the-art methodology proposed by Basireddy et al. [5]. In [5], the researchers have proposed a workload management system, which classifies workloads of the executing applications based on Memory Reads Per Instruction (MRPI) metric and manages DVFS levels of cores based on it.

RSA encryption and decryption was performed for 512, 1024, 2048, 4096 bits for 10 secs for each types. Based on the TSMP values in Table 1 DATE is 4.30% more secure for RSA encryption and decryption than the Linux Performance Governor. Whereas, DATE is 139.24% more secure for Streamcluster benchmark against temperature based side-channel attacks than the Linux Performance Governor. DATE is also 35.46% more secure than MRPI [5] for the Streamcluster benchmark.

TABLE 1: TSMP values for different methodologies

| App | OnDemand | Performance | MRPI | DATE |
|---|---|---|---|---|
| *Blackscholes* | 0.001968 | 0.001124 | 0.001792 | 0.003496 |
| *Streamcluster* | 0.000479 | 0.000450 | 0.000846 | 0.001146 |
| *RSA* | 0.001811 | 0.002439 | 0.001912 | 0.002544 |
| *Youtube* | 0.001661 | 0.002840 | 0.002906 | 0.003012 |

TABLE 2: Thermal cycle reduction comparison with DATE

| App | OnDemand | Performance | MRPI |
|---|---|---|---|
| *Blackscholes* | 40.57% | 67.42% | 46.96% |
| *Streamcluster* | 59.59% | 62.21% | 28.45% |
| *RSA* | 29.34% | -2.91% | 23.54% |
| *Youtube* | 48.15% | 2.76% | 1.06% |

Fig. 17 graphically shows the temperature peaks achieved for the Blackscholes application using DATE, which closely resonates with *baseline temperature* (see Fig. 1). In comparison, Fig. 16 highlights the temperature peaks achieved for the Blackscholes application while executing on ondemand and performance governors of Linux. From the figures (Fig. 16 and Fig. 17) it is also very evident that DATE is able to achieve the reduction in spatial thermal gradient as well as temporal thermal gradient while the overall operating temperature of the CPUs is also reduced at the same time (also see the histogram comparison in Fig. 18).

**Note**: For data dependent applications over a network such as playing a video on Youtube in Chromium browser, it was still difficult to reduce the $Spatial\ Themal\ Deviation$ due to data dependencies over the Internet or due to workload imbalance between the CPU cores.

### 5.3 Effect on Device Reliability

Thermal gradient (spatial and temporal) and thermal cycling [42] play important role in the reliability of the system in terms of lifespan. The rate of temperature changes on the system affects the lifespan of the device over time due to degradation in the structural integrity and chemical property. More importantly, thermal cycle reduces the whole systems MTTF (Mean-time-to-failure) as the amplitude and/or frequency of thermal cycles increase. The number of the thermal cycles ($N_{TC}(i)$) that can result in the occurrence of failure due to the $i^{th}$ thermal cycle can be deduced from the modified Coffin-Manson equation (see Eq. 12) mentioned in [43].

$$N_{TC}(i) = A_{TC}(\delta T_i - T_{th})^{-b} exp(\frac{E_{aTC}}{T_{max_i}}) \tag{12}$$

In Eq. 12, $\delta T_i$ represents the maximum thermal amplitude change of the $i^{th}$ thermal cycle, $T_{th}$ is the threshold temperature of the component/device at which inelastic deformation begins, $A_{TC}$ is an empirically determined constant [43], $b$ is the Coffin-Manson exponent constant, $E_{aTC}$ is the activation energy and $T_{max_i}$ is the maximum temperature in the $i^{th}$ thermal cycle. Therefore, from Eq. 12 we can deduce the MTTF ($MTTF_{TC}$) [44] related to thermal cycle ($N_{TC}$) as follows:

$$MTTF_{TC} = \frac{N_{TC} \sum_{i=0}^{n} t_i}{n} \tag{13}$$

In Eq. 13, $t_i$ is the temperature of the component/device at $i^{th}$ thermal cycle, and $n$ is the total number (frequency) of thermal cycle. Since, we focused most of the experiments on the Odroid XU4 MPSoC, and if we assume that the number of the thermal cycles ($N_{TC}$) that can result in the occurrence of failure due to the $i^{th}$ thermal cycle is same for the same device platform, then we can easily evaluate the $MTTF_{TC}$ for each executing application on individual thermal management methodologies. Now, if we take the example of Blackscholes application being executed on the MPSoC using Linux's ondemand governor and assume that $t_i$ is the baseline maximum thermal deviation i.e. $t_i = \tau$ since baseline temperature is the same during all the experiments performed during the time of experimentation, then from Fig. 13 and Fig. 15 we could notice that $t_i$ ($\tau$) is 20 and $n$ (frequency of thermal cycle) is 456. Hence, putting the value of $t_i$ and $n$ in Eq. 13, we get the value of $MTTF_{TC}$ as $\frac{N_{TC} \times 9140}{456}$. Now, if we consider the DATE methodology, where Blackscholes is executed on the Odroid XU4 using DATE, $t_i$ is 7 (see Fig. 13) and $n$ is 271 (see Fig. 15), then we get the value of $MTTF_{TC}$ as $\frac{N_{TC} \times 1904}{271}$. Therefore, comparing the $MTTF_{TC}$ of executing Blackscholes using the Linux's ondemand governor and the DATE, the DATE methodology is 2.85 (approx.) ($\simeq \frac{N_{TC} \times 9140}{456} / \frac{N_{TC} \times 1904}{271}$) times more reliable and improve the lifespan of the device over time.

### 5.4 Discussion and Future Direction

For applications with data dependencies over the Internet such as Youtube or applications with workload imbalance between the CPU cores, a possible solution to reduce the Spatial Themal Deviation is to add noise using software/algorithmic implementation to the temperature variance such that it becomes very difficult to understand the pattern of execution or data being processed on the CPU cores. Here, noise could be regarded as another random execution of secondary tasks in order to keep the CPU temperature high so that temperature dissipated due to execution of the current task could be masked. The main challenge in using such a solution is that to predict when to add the noise in order to mask thermal dissipation of the currently executing task properly.

## 6 Related Work

We have also chosen the state-of-the-art approaches [16]–[18], which is closely related to our study and address the temperature side-channel attacks on general purpose multiprocessor/core devices but not on embedded platforms, and compared the methodologies with the DATE methodology. In the study by Gu et al. [16], (referred to as *3D-TASCS*), the scholars have proposed a thermal-aware hardware functional unit that introduces noise to mask activity of encryption and decryption in order to secure against temperature based side-channel attack. Whereas, in the study by Long et al. [17], the researchers have proposed an approach to apply multiple transmitter to one receiver by generating noise to mask temperature behavior on the main transmitter and also establishing a new communication protocol between the transmitter and receiver. In [17], a transmitter is the CPU core, which is executing the main task or set of tasks such as encryption/decryption, whereas, a receiver is the CPU core, which is establishing the thermal side-channel attack. In the study by Knechtel et al. [18], referred to as *3D-IC*, the researchers exploit the specifics of material and structural properties in 3D ICs during design time exploration, thereby decorrelating the thermal behavior from underlying power and activity patterns.

| | | Blackscholes | | | | Streamcluster | | | | RSA | | | | Youtube | | | |
|---|---|---|---|---|---|---|---|---|---|---|---|---|---|---|---|---|---|
| | Base. T. | OnDemand | Performance | MRPI | DATE | OnDemand | Performance | MRPI | DATE | OnDemand | Performance | MRPI | DATE | OnDemand | Performance | MRPI | DATE |
| CPU 4 | 61 | 76 | 78 | 76 | 68 | 95 | 95 | 90 | 90 | 83 | 86 | 84 | 74 | 91 | 90 | 91 | 84 |
| CPU 5 | 64 | 75 | 75 | 76 | 65 | 88 | 92 | 93 | 86 | 78 | 75 | 84 | 82 | 87 | 86 | 86 | 80 |
| CPU 6 | 68 | 84 | 85 | 84 | 72 | 96 | 95 | 96 | 90 | 86 | 86 | 77 | 82 | 96 | 90 | 95 | 90 |
| CPU 7 | 62 | 82 | 82 | 72 | 67 | 95 | 95 | 95 | 92 | 89 | 81 | 80 | 77 | 95 | 94 | 95 | 88 |
| Base. T. Diff | CPU 4 | 15 | 17 | 15 | **7** | **34** | **34** | 29 | 29 | 22 | **25** | **23** | 13 | 30 | 29 | 30 | 23 |
| | CPU 5 | 11 | 11 | 12 | 1 | 24 | 28 | 29 | 22 | 14 | 11 | 20 | **18** | 23 | 24 | 22 | 16 |
| | CPU 6 | 16 | 17 | **16** | 4 | 28 | 27 | 28 | 22 | 18 | 18 | 9 | 14 | 28 | 22 | 27 | 22 |
| | CPU 7 | **20** | **20** | 10 | 5 | 33 | 33 | **33** | **30** | **27** | 19 | 18 | 15 | **33** | **32** | **33** | **26** |
| τ | | **20** | **20** | **16** | **7** | **34** | **34** | **33** | **30** | **27** | **25** | **23** | **18** | **33** | **32** | **33** | **26** |

Fig. 13: Table showing maximum operating temperature of ARM A-15 (big) CPUs for different applications along with *baseline temperature* and values of *Baseline Maximum Thermal Deviation* ($\tau$) for corresponding CPUs

| | Blackscholes | | | | Streamcluster | | | | RSA | | | | Youtube | | | |
|---|---|---|---|---|---|---|---|---|---|---|---|---|---|---|---|---|
| | OnDemand | Performance | MRPI | DATE | OnDemand | Performance | MRPI | DATE | OnDemand | Performance | MRPI | DATE | OnDemand | Performance | MRPI | DATE |
| CPU 4 | 60 | 62 | 62 | 61 | 62 | 64 | 64 | 65 | 64 | 56 | 62 | 60 | 69 | 59 | 63 | 59 |
| CPU 5 | 61 | 63 | 63 | 61 | 63 | 65 | 67 | 64 | 65 | 57 | 65 | 61 | 66 | 60 | 63 | 60 |
| CPU 6 | 64 | 65 | 65 | 64 | 66 | 67 | 65 | 69 | 68 | 59 | 65 | 67 | 70 | 63 | 65 | 65 |
| CPU 7 | 61 | 62 | 62 | 62 | 62 | 64 | 64 | 67 | 64 | 56 | 64 | 63 | 67 | 59 | 62 | 63 |
| ω | **21** | **20** | **19** | **8** | **33** | **31** | **31** | **25** | **25** | **30** | **22** | **21** | **28** | **35** | **33** | **25** |

Fig. 14: Table showing least average operating temperature of ARM A-15 (big) CPUs for different applications and values of *Spatial Maximum Thermal Deviation* ($\tau$) for corresponding CPUs

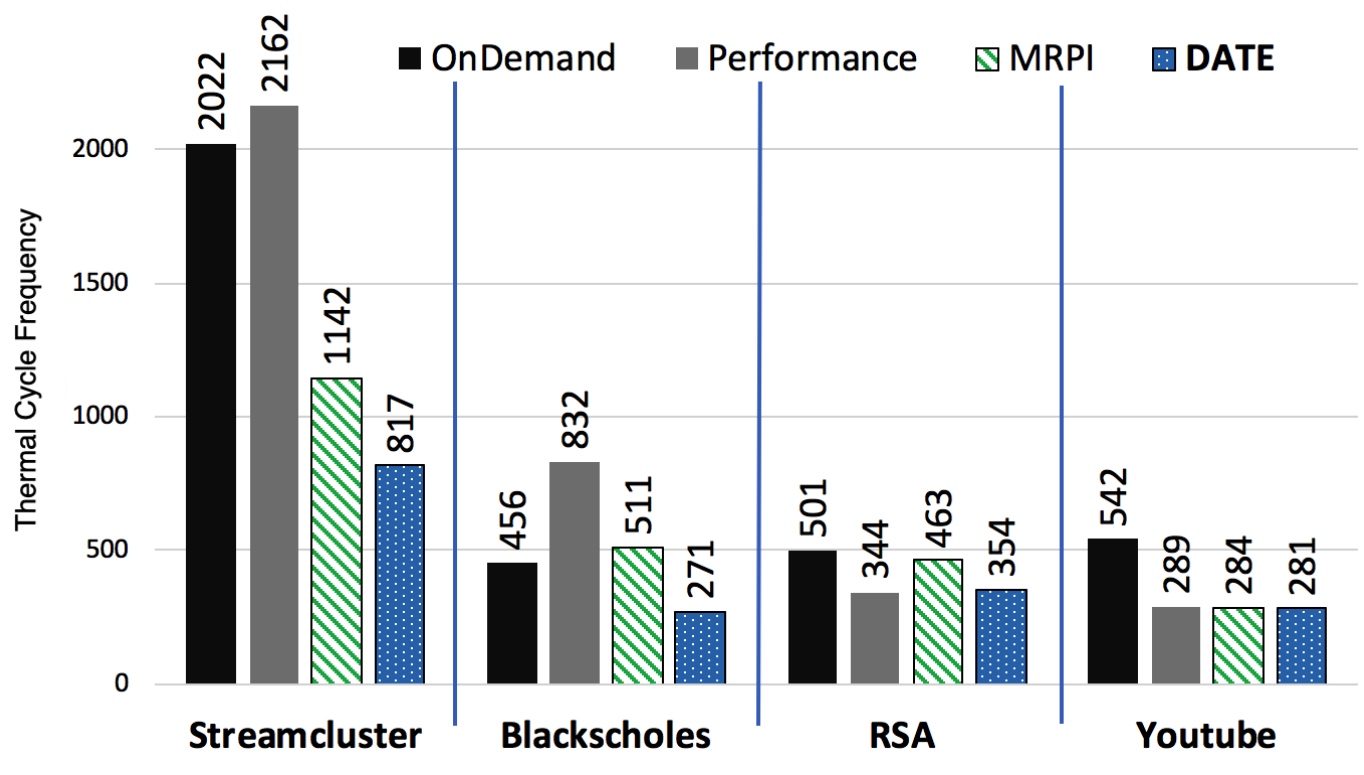

Fig. 15: Frequency of thermal cycles for different benchmarking applications on different power and mapping schemes

TABLE 3: Comparative study of $3D-TASCS$ [16], *Long et al.* [17] and $3D-IC$ [18] with *DATE*

| Method | Sim | S.Int. | H.Int. | Easy | STR | TTR | T.C. |
|---|---|---|---|---|---|---|---|
| $3D-TASCS$ [16] | ✓ | ✗ | ✓ | ✗ | ✓ | ✓ | ✗ |
| *Long et al.* [17] | ✓ | ✓ | ✗ | ✓ | ✗ | ✗ | ✗ |
| $3D-IC$ [18] | ✓ | ✗ | ✓ | ✗ | ✗ | ✗ | ✗ |
| ***DATE*** | ✗ | ✓ | ✗ | ✓ | ✓ | ✓ | ✓ |

We performed comparison based on the following criterion:

- Whether the methodology is implemented on a simulator or not implemented on a real hardware platform and it is denoted as **Sim** (Simulation) in this research.
- Whether the methodology is integrated through application/software/algorithmically and it is denoted as **S.Int.** (Software Integration).
- Whether the methodology is integrated through hardware or by developing special integrated circuit modifications and it is denoted as **H. Int.** (Hardware Integration).
- Whether the methodology is able to be deployed with ease on existing embedded platforms utilizing DVFS enabled multi-processors and it is denoted as **Easy**.
- Whether the methodology affects the spatial thermal gradient of the processing elements actively and it is denoted as **STR** (Spatial Thermal Reduction).
- Whether the methodology affects the temporal thermal gradient of the processing elements actively and it is denoted as **TTR** (Temporal Thermal Reduction).
- Whether the methodology affects the thermal cycle frequency or thermal cycling of the processing elements of the embedded MPSoC actively and it is denoted as **T.C.** (Thermal Cycles).

Table 3 shows the outcome of the comparison, where it could be clearly noticed that DATE is not just easy to implement but could be deployed easily on existing hardware as well as hardware systems, utilizing DVFS enabled multi-processors, to come in the future through algorithmic implementation and would not just secure the system from temperature based side-channel attack but also affect the thermal gradient of the operating CPU cores actively, hence, affecting reliability of the device as well. Another important observation is that because the DATE was implemented on a real hardware platform, the methodology provides more confidence in the research work being pursued since simulation platform could differ from the practical results achieved from real devices.

## 7 Conclusion

In this paper, we proposed a universal metric to quantify security of embedded devices against temperature side-channel attacks and a novel approach, DATE, to secure DVFS enabled MPSoCs against the same. We also performed a real thermal side channel attack on Samsung Note 9 mobile device utilizing machine learning to release the feasibility of such an attack. Experimental studies evaluated on the Exynos 5422 MPSoC and comparative study with the state-of-the-art proves the efficacy of DATE in securing the device from thermal side-channel attack. Therefore, from experimental evaluation we also proved that DATE is not just effective against temperature side-channel attacks but is also effective to enhance reliability in terms of the lifespan of the device.

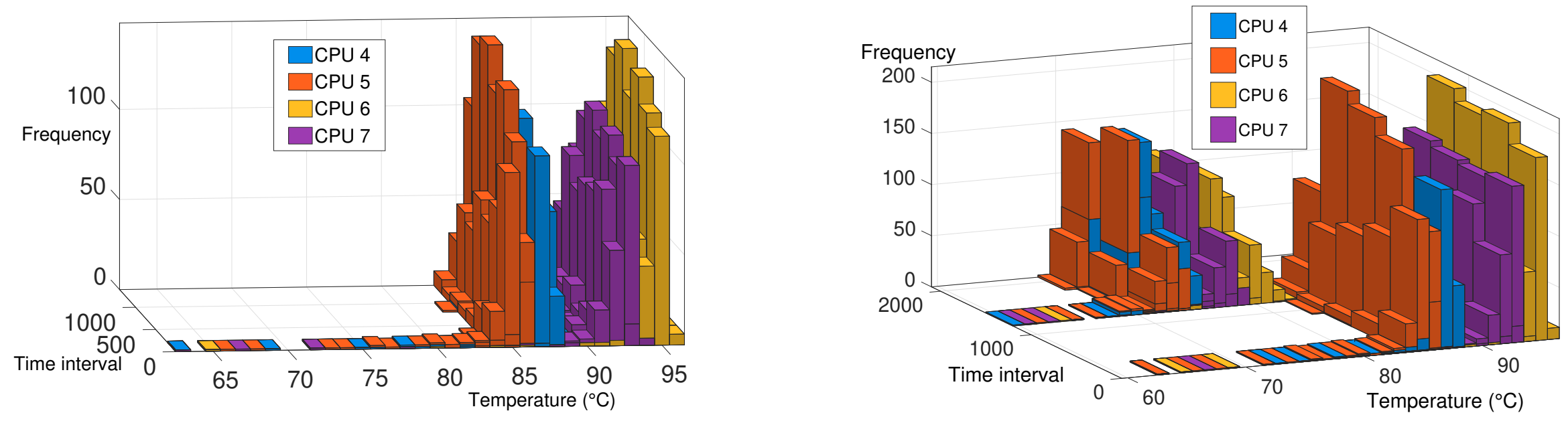

(a) Using Ondemand governor (b) Using Performance governor

Fig. 16: Temperature peaks of 4 ARM A-15 big CPU cores while executing Blackscholes benchmark using Linux's Ondemand and Performance governors

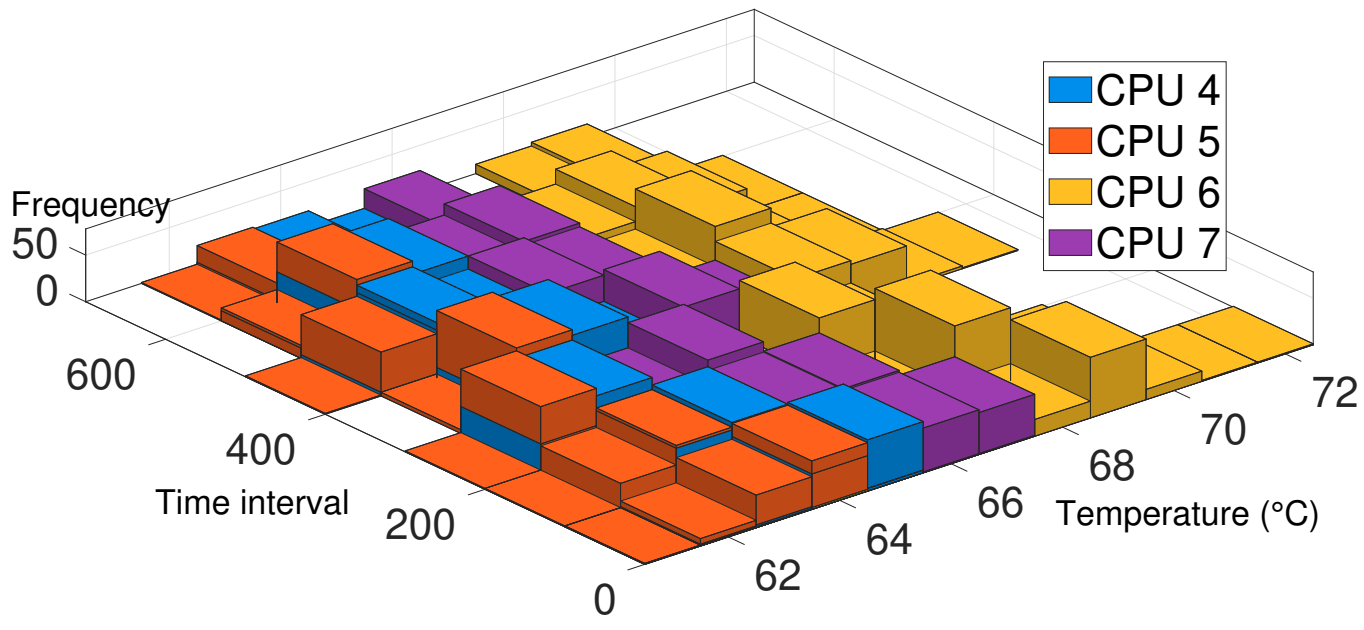

Fig. 17: Temperature peaks of 4 ARM A-15 big CPU cores while executing Blackscholes benchmark using DATE

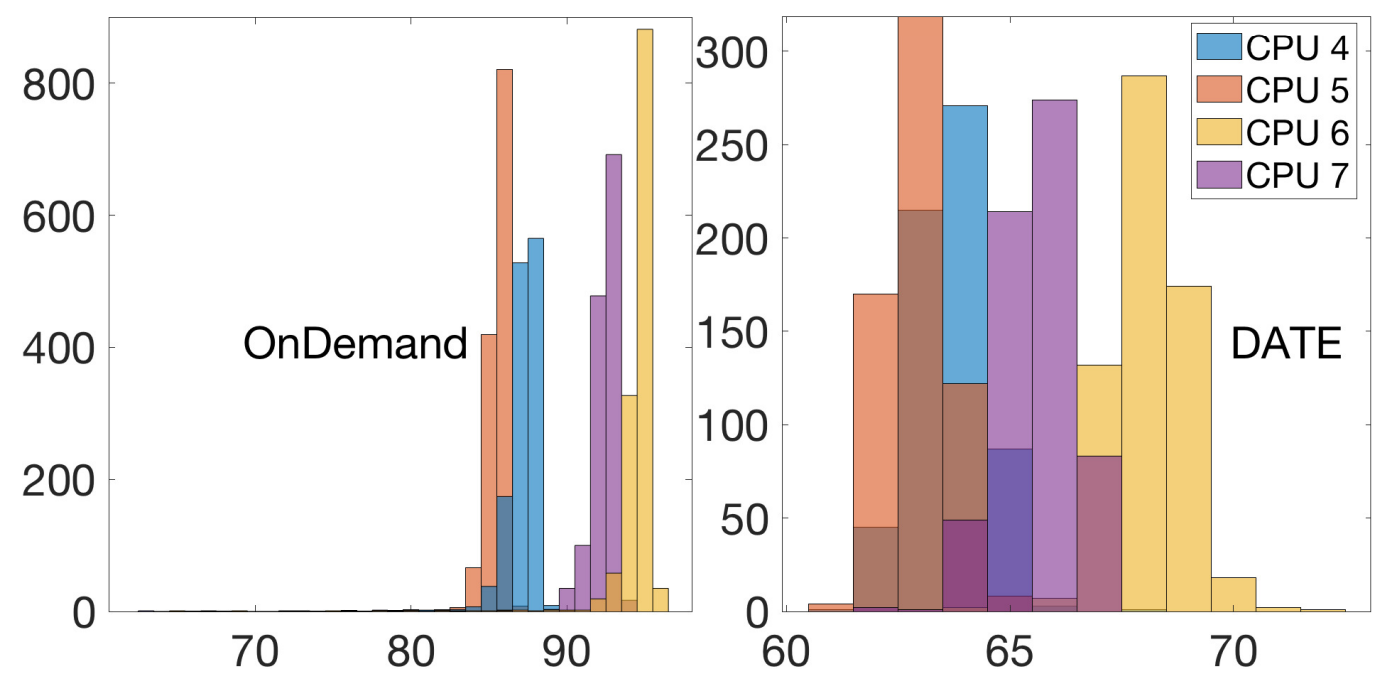

Fig. 18: Histograms of temperature peaks of ARM big CPUs for Blackscholes using ondemand governor vs DATE (Frequency vs temperature (°C))